\documentclass[aps,prd,showkeys,showpacs,twocolumn]{revtex4}
\usepackage[english]{babel}
\usepackage{amsmath,graphicx}
\usepackage{palatino}
\usepackage[colorinlistoftodos]{todonotes}
\usepackage{changes}
\usepackage[colorlinks=true,
            linkcolor=blue,
            urlcolor=blue,
            citecolor=blue]{hyperref}
\newcommand{\assign}{:=}
\newcommand{\nosymbol}{}

\newcommand{\tmop}[1]{\ensuremath{\operatorname{#1}}}

\begin{document}


\title{Deflection angle of photon through dark matter by black holes and wormholes using the Gauss-Bonnet theorem }

\author{A. {\"O}vg{\"u}n}
\email{ali.ovgun@pucv.cl}
\affiliation{Instituto de F{\'i}sica, Pontificia Universidad Cat{\'o}lica de
Valpara{\'i}so, Casilla 4950, Valpara{\'i}so, Chile}
\affiliation{Physics Department, Arts and Sciences Faculty, Eastern Mediterranean University, Famagusta, 99628 North Cyprus via Mersin 10, Turkey}

\date{\today}

\begin{abstract}
 In this research, we use the Gibbons-Werner method (Gauss-Bonnet theorem) on the optical geometry of a black hole and wormhole, extending the calculation of the weak gravitational lensing within the Maxwell's fish eye-like profile and dark matter medium. The angle is seen as a partially topological effect and the Gibbons-Werner method can be used on any asymptotically flat Riemannian optical geometry of compact objects in dark matter medium.
   
  \end{abstract}



{\maketitle}
\section{Introduction}
Black holes are an essential component of our universe, and one of the most important discoveries in astrophysics is that when stars die, they can collapse to become extremely small objects. Black holes provide an important opportunity for probing and testing the fundamental laws of the universe. For example, gravitational waves from black holes and neutron star mergers have been recently detected \cite{Abbott:2016blz}. Black holes may also hint at the nature of quantum gravity at small scales that change the area law of entropy.  Quantum gravity is far from understood, though theoretically it has seen tremendous progress and in a few years the Event Horizon Telescope may provide some more information about it  \cite{Giddings:2016btb,Barrau:2017rwl,nutku,Akiyama:2019cqa}. 

 In 1854, Maxwell presented the solution to a mathematical problem related to the passage of rays through a sphere of variable refractive index, and he noted that the potential existence of a medium of this kind would possess exceptional optical properties \cite{Maxwell}. This is similar to the reflection of the crystalline lens in fish. This optical tool is Maxwell's fish-eye (MFE),  the condition in which all light rays form circular trajectories. It was a remarkable accomplishment  to visualize that a lens whose refractive index increases towards a point could form perfect images \cite{pendy}.

Luneburg discovered that the ray propagation of MFE is equivalent to ray propagation on a homogeneous sphere with a unit radius and a unit refractive index within geometrical optics \cite{lune}. This showed that the imaging of variants that have been applied to microwave devices and the fish-eye lens in photography, form an extremely wide-angled image, almost hemispherical in coverage.  In 2009, Leonhardt showed that MFE is also good for waves, and it enables the production of super-resolution imaging with perfect lensing, which requires negative refractive-index materials. This began a debate and offered a rich area of research to explore \cite{leon, leonn,leonnn}.It was shown that perfect imaging in a homogeneous 3-dimensional region is also possible \cite{3dmfe}. MFE happens when all light rays, arising from any point within, converge at its conjugate, which means that power released from a source can only be fully absorbed at its image point, resulting in perfect imaging. There has been a rapid increase in the importance of perfect imaging in theoretical and experimental optics \cite{tomas,yang,gun}.

Fermat’s principle says that light rays always follow extremal optical paths with a path length being measured by the refractive index $n$. The formula for MFE indicates the interesting possibility that rays generate a perfect image in a black hole. The refractive index depends only on the distance $r$ from the origin \cite{tomas,yang}. In this paper, we try to understand the effect of MFE-like profile on the deflection angle. For simplicity, we use the uniform MFE-like profile, which is different from a non-uniform MFE profile.

Gravitational lensing is a useful tool of astronomy and astrophysics  \cite{lens}, in which the light rays from distant stars and galaxies are deflected by a planet, a black hole, or dark matter  \cite{Bozza:2009yw,Stefanov:2010xz}. The detection of dark matter filaments \cite{dark} using weak deflection is a very relevant topic because it can help in understanding the large-scale structure of the universe \cite{Bartelmann:1999yn}.
To build a sky map (the index of refractive of the entire visible universe), there is ongoing research on the observation of the effect of cosmological weak deflection on temperature fluctuations in the cosmic microwave background (CMB) \cite{CMB2}. From a theoretical point of view, new methods have been proposed to calculated the deflection angle. One of them is the Gauss-Bonnet theorem (GBT), which was first proposed by Gibbons and Werner, using optical geometry \cite{Gibbons:2008rj,Gibbons:2008hb,Werner:2012rc}. The deflection angle is seen as a partially topological effect that can be calculated by integrating the Gaussian curvature of the optical metric outwards from the light ray using the following equation: \cite{Gibbons:2008rj,Gibbons:2008hb}
\begin{eqnarray}
  \hat{\alpha} = - \int\int_{ D_{\infty}} K d A. 
  \label{int0}
\end{eqnarray}
Since the Gibbons and Werner's paper on weak deflection angle by GBT provided a unique perspective, this method has been applied in various cases \cite{Ishihara:2016vdc,Sakalli:2017ewb,Jusufi:2017lsl,Ono:2017pie,Jusufi:2017hed,Ishihara:2016sfv,Jusufi:2017vta,Arakida:2017hrm,Jusufi:2017mav,Ovgun:2018tua,Ovgun:2018fte,Ovgun:2018prw,Jusufi:2018jof,Ovgun:2018xys,Ovgun:2018ran,Jusufi:2017uhh,Javed:2019qyg,Ovgun:2018fnk,Ono:2018ybw,Ono:2018jrv,Crisnejo:2018uyn,Crisnejo:2018ppm,Asada:2017vxl}. 

Dark matter is making up 27\% of the total mass-energy of the universe \cite{dm1}. We can only detect dark matter from its gravitational interactions and we only know that dark matter is non-baryonic, non-relativistic (or cold), and it has weak non-gravitational interactions. There are many dark matter candidates, such as weakly interacting massive particles (WIMPs), super-WIMPs, axions, and sterile neutrinos \cite{dm2}. It has been proposed that dark matter is a composite, such as the dark atom model, which we investigate here using the deflection of light through it. Dark matter, although suppressed, generally has electromagnetic interactions \cite{Latimer:2013rja} such that the medium of dark matter should have some optical properties that a traveling photon can sense because of the frequency-dependent refractive index. The refractive index regulates the speed at which a wave is propagated via a medium. The particles of dark matter do not get electrically charged, but they can couple to other particles that have a virtual electromagnetic charge and can also couple to photons  \cite{
Latimer:2017lwm,Latimer:2016kdg,Kvam:2016plz,Whitcomb:2017iiq}. To find the amplitude of dark matter annihilation into two photons, we must first calculate the scattering amplitude. One can obtain the index of the refractive of light, where the real part is related to the speed of propagation.

To investigate weak deflection through dark matter, we consider the propagation effects for the case that particles of dark matter (warm thermal relics or axion-like particles) have low mass  whose number density is larger than ordinary matter. Put simply, dark matter interacts with photons (if only through quantum fluctuations), resulting in a refractive index. The relationship between the refractive index and the forward Compton amplitude at relatively low photon energies \cite{Latimer:2013rja} ($
\mathcal { M } _ { \mathrm { fwd } } \sim - \varepsilon ^ { 2 } e ^ { 2 }
$) is 

\begin{equation}
    n = 1 + \frac { \rho } { 4 m _ { d m } ^ { 2 } \omega ^ { 2 } } \mathcal { M } _ { \mathrm { fwd } },
\end{equation}
where $\omega$ is the measured photon frequency, and $\rho = 1.1 \times 10^{-6} GeV/cm^3$ is the present day dark matter density \cite{Latimer:2013rja}. Neglecting spin, the amplitude will be a real and even function of $\omega$ (for photon energies below the inelastic threshold); additionally, the coefficients of the $O(\omega^{2n})$ terms are positive and the spin dependent interactions can lead to odd powers in the expansion about $\omega$. Their presence could give information on the spin of dark matter. Hence, the refractive index becomes: \cite{Latimer:2013rja}

\begin{equation}
  n = 1 + \frac { \rho } { 4 m _ { d m } ^ { 2 } } \left[ \frac { A } { \omega ^ { 2 } } + B + C \omega ^ { 2 } + \mathcal { O } \left( \omega ^ { 4 } \right) \right].
\end{equation}
To do so, we suppose that the photons can be deflected through the dark matter due to the dispersive effects. We use the index of refractive $n(\omega)$ that is manipulated by the scattering amplitude of the light and dark-matter in the forward \cite{Latimer:2013rja}.
 
  Gravitational lensing in plasma has been studied in various cases \cite{Crisnejo:2018uyn,Crisnejo:2018ppm,Tsupko:2013cqa,BisnovatyiKogan:2010ar,Bisnovatyi-Kogan:2015dxa}. For the first time, Bisnovatyi-Kogan and Latimer showed that due to dispersive properties of plasma even in the homogeneous plasma the gravitational deflection differs from
vacuum deflection angle \cite{BisnovatyiKogan:2010ar,Bisnovatyi-Kogan:2015dxa}. Moreover, it was shown that deflection angle is increased due to the presence of plasma \cite{Tsupko:2013cqa}. Afterwards, Crisnejo and Gallo calculated weak lensing in a plasma medium
using the GBT \cite{Crisnejo:2018uyn}.

The main motivation of this research is to shed light on the unexpected features of spacetimes in regards to MFE-like profile and to derive the deflection angle of black holes using the Gauss-Bonnet theorem in weak limits for a dark matter medium. We suppose that the refractive index of the medium is spatially non-uniform but it is uniform at large distances. We also investigate the effect of various parameters on the refractive index of the medium, which has not been covered in previous studies.

\section{Effect of medium on deflection angle of Schwarzschild black hole using the Gauss-Bonnet theorem }

In this section, we shall first describe the black hole solution in a
static and spherically symmetric spacetime. Then we apply the MFE-like profile within the GBT to calculate the weak deflection angle.

The Schwarzschild black hole spacetime reads
\begin{equation}
  ds^2 = - f (r) dt^2 + g (r)^{\nosymbol} dr^2 + r^2  (d \theta^2 + \sin^2
  \theta d \varphi^2), \label{2}
\end{equation}
with the metric functions
\begin{eqnarray}
  f (r) & = g (r)^{- 1} = & 1 - \frac{2 M}{r} . 
\end{eqnarray}
Analysis of the geodesics equation, the ray equation is calculated by
\begin{equation} \varphi = \int \frac{b \sqrt{g (r)} \tmop{dr}}{r^2  \sqrt{\frac{1}{f (r)} -
   \frac{b^2}{r^2}}}, \end{equation}
where $b$ is the impact parameter of the unperturbed photon.

Our universe is homogeneous and isotropic on large scales. Now we
consider isotropic coordinates which are non-singular at the horizon and the
time direction is a Killing vector. Moreover, time slices become Euclidean
with a conformal factor and one can calculate the index of refractive, $n$, of
light rays around the black hole. Other important feature of the isotropic
coordinates is that they satisfy Landau's condition of the coordinate clock
synchronization
\begin{equation} \frac{\partial}{\partial x_j}  \left( - \frac{g_{0 i}}{g_{00}} \right) =
   \frac{\partial}{\partial x_i}  \left( - \frac{g_{0 j}}{g_{00}} \right)  (i,
   j = 1, 2, 3) . \end{equation}
Using the following transformation
\begin{equation}
  r = \rho \left( 1 + \frac{M}{2 \rho} \right)^2, \label{21}
\end{equation}
the Schwarzschild black hole rewritten in isotropic coordinates (where $\rho$
is isotropic radial coordinate) \cite{Crisnejo:2018uyn}
\begin{equation}
  ds^2 = - F (\rho) dt^2 + G (\rho)  (d \rho^2 + \rho^2 d \Omega^2),
  \label{22}
\end{equation}
with
\begin{equation}
  F (\rho) = \left( \frac{\rho - \frac{M}{2}}{\rho + \frac{M}{2}} \right)^2,
  \label{23} \tmop{and} G(\rho) = \left( \frac{\rho + \frac{M}{2}}{\rho}
  \right)^4 .
\end{equation}
The metric becomes nonsingular at the horizon $r = 2 M$. It can be also
written in Fermat form of metric:
\begin{equation}
  ds^2 = F (\rho) [- dt^2 + n (\rho)^2  (d \rho^2 + \rho^2 d \Omega^2)],
  \label{25}
\end{equation} with the index of
refractive $n (\rho) = \frac{c}{v (\rho)}$. For
the Schwarzschild black hole medium, the refractive index reads
\begin{equation}
  n = \frac{\left( 1_{\nosymbol} + \frac{M}{2\rho} \right)^3}{\left(
  1_{\nosymbol} - \frac{M}{2\rho} \right)}, \label{11}
\end{equation}
and it can be approximated for large $\rho \gg M$
\begin{equation} n \approx 1_{\nosymbol} + \frac{2 M}{\rho} . \label{12} \end{equation}
Now the ray equation becomes
\begin{equation} \varphi = \int \frac{bd \rho}{\rho^2  \sqrt{n^2 - \frac{b^2}{\rho^2}}} . \end{equation}
To discuss the deflection angle and extract information of MFE-like profile, the GBT will be used instead of the null geodesics
method. The GBT is calculated using the negative Gauss curvature of the
optical metric.

\subsection{Case 1:} Let us start from the constant case for the medium $n(r)$ as refractive
index:
\begin{equation} n_{m} =  n_0, \end{equation}
where $n_0$ is a constant refractive index of the medium,
and here we consider the GBT to obtain the deflection angle in a medium
in weak field limits. 

Let us write the optical Schwarzschild spacetime in an
equatorial plane \cite{Crisnejo:2018uyn}:
\begin{equation}
  d\sigma^2 =  \frac{n_{m}^2}{f(\rho)}[g (\rho) dr^2 + \rho^2 d \varphi^2]. \label{2222}
\end{equation}

Then we calculate the Gaussian optical curvature
\begin{equation}
K=-2\,{\frac {M}{{{\it n_0}}^{2}{\rho}^{3}}}+3\,{\frac {{M}^{2}}{{{\it n_0}}^
{2}{\rho}^{4}}}+O(M^3),
\end{equation}
which is everywhere negative that gives a universal property of black hole metrics \cite{Gibbons:2008hb}.

It reduces to this form at linear order of $M$:
\begin{equation}
K\approx -2\,{\frac {M}{{{\it n_0}}^{2}{\rho}^{3}}}.\end{equation}

This result will be used to evaluate the deflection angle using a non-singular
domain outside the light ray ( $D_\rho$, with boundary $\partial
D_\rho = \gamma \cup C_\rho$) \cite{Gibbons:2008rj}:
\begin{equation}
  \iint_{D_\rho} K \hspace{0.17em} d S + \oint_{\partial
  D_\rho} \kappa \hspace{0.17em} d t + \sum_i \theta_i = 2 \pi
  \chi (D_\rho),
\end{equation}
where $\kappa$ stands for the geodesic curvature and $K$
is \ Gaussian optical curvature, with the exterior angles
$\theta_i$=($\theta_{O}$, $\theta_{S}$) \ and the Euler
characteristic number $\chi (D_\rho) = 1$. At weak limits, ( $\rho
\rightarrow \infty$), \ $\theta_{O} + \theta_{S} \rightarrow
\pi$. Then the GBT reduces to
\begin{equation}
  \iint_{D_\rho} K \hspace{0.17em} d S + \oint_{C_\rho} \kappa
  \hspace{0.17em} d t \overset{\rho \rightarrow \infty}{=}
  \iint_{D_{\infty}} K \hspace{0.17em} d S + \int_0^{\pi +
  \hat{\alpha}} d \varphi = \pi .
\end{equation}
For the geodesics $\gamma$, \ geodesic curvature
vanishes $\kappa (\gamma) = 0$, and we have
\begin{equation}
  \kappa (C_\rho) = | \nabla_{\dot{C}_\rho}  \dot{C}_\rho |,
\end{equation}
with $C_\rho \assign \rho (\varphi) = \rho =$constant. The GBT becomes
\begin{equation}
\lim_{\rho\rightarrow\infty}\int_{0}^{\pi+\hat{\alpha}}\left[\kappa\frac{d\sigma}{d\varphi}\right]\bigg|_{C_{\rho}}d\varphi    =\pi-\lim_{\rightarrow \infty }\int \int_{D_{\rho}} K dS,
\label{alpha-bonnet}
\end{equation}

and one can calculate
\begin{equation}
\frac{d\sigma}{d\varphi}\bigg|_{C_{\rho}} =n_{m}\bigg(\frac{r^3}{\rho-2M}\bigg)^{1/2},
\end{equation}
where for very large radial distance
\begin{equation}
\ \kappa (C_\rho) d t = d \hspace{0.17em} \varphi.
\end{equation}

Therefore, as expected for this number density profile and
physical metric (which imply that the optical metric is asymptotically
Euclidean) we corroborate that 
\begin{equation}
\lim_{\rho\rightarrow\infty}\kappa\frac{d\sigma}{d\varphi}\bigg|_{C_{\rho}}=1.\label{khomons}
\end{equation}
At linear order in $M$, it follows using \eqref{alpha-bonnet} in
the limit $\rho\rightarrow\infty$, and taking the geodesic curve $\gamma$
approximated by its flat Euclidean version parametrized as $\rho=b/\sin\varphi$,
with $b$ representing the im pact parameter in the physical spacetime
that 
\begin{equation}
\hat{\alpha}=-\lim_{\rho\rightarrow\infty}\int_{0}^{\pi}\int_{\frac{b}{\sin\varphi}}^{\rho}K dS.\label{alpha1}
\end{equation}
After nontrivial calculation, we calculate the deflection angle of the
Schwarzschild black hole in medium for the leading order terms is
\begin{equation}
\hat{\alpha}=4\,{\frac {M}{{\it n_0}\,b}},\label{phs}
\end{equation}
which agrees with the well-known results in the
limit at which its presence is negligible ($n_0=1$)
this expression reduces to the known vacuum formula $\hat{\alpha}=4\,{\frac{M}{b}}.$
So that GBT exhibits a partially topological effect. This
method can be used in any asymptotically flat Riemannian optical metrics.

\subsection{Case 2:}

Now we apply the different model of MFE-like medium \cite{leonn}
\begin{equation}
n = \frac{z_0}{1 + z^2}, \end{equation}
where $z_0$ and $z$ are a constant.

The Gaussian curvature of the optical
metric approximating in leading orders is everywhere
negative and found as :

\begin{equation}
K=-2\,{\frac { \left( {z}^{2}+1 \right) ^{2}M}{{{\it z_0}}^{2}{\rho}^{3}}}+O(M^3),
\end{equation}

Then using the same method we calculate the  deflection angle as follows:
\begin{equation}
  \hat{\alpha} \simeq 4\,{\frac {M{z}^{2}}{{\it z_0}\,b}}+4\,{\frac {M}{{\it z_0}\,b}} \hspace{0.17em}.
\end{equation}
At the $z =0$ and $z_0=1$, it reduces to exact Schwarzschild case.

\subsection{Case 3:}

The refractive index for the dark matter medium  \cite{Latimer:2013rja}

\begin{equation}n ( \omega ) =1+\beta A_0+A_2 \omega^2 \label{dm} \end{equation}

where $\beta=\frac { \rho_0 } { 4 m ^ { 2 } \omega ^ { 2 } } $ and $\rho_0$ the mass density of the scattered dark matter particles and $A _ { 0 } = - 2 \varepsilon ^ { 2 } e ^ { 2 } \text { and } A _ { 2 j } \geq 0
$.  

The terms in $
\mathcal { O } \left( \omega ^ { 2 } \right)
$ and higher are related to the polarizability of the dark-matter candidate. 
Note that, the order of $\omega^{-2}$ is due to the charged dark matter candidate and $\omega^{2}$ for a neutral dark matter candidate. Moreover, there may be a linear term in $\omega$ when parity and charge-parity asymmetries are present.

The Gaussian curvature is obtained as:
\begin{equation}
K \approx -2\,{\frac {M}{ \left( {\it A_2}\,{\omega}^{2}+\beta\,{\it 
A_0}+1 \right) ^{2}{\rho}^{3}}}+O(M^3)\end{equation}
The deflection angle is found as follows:

\begin{equation}
\hat{\alpha}= 4\,{\frac {M}{ \left( {\it A_2}\,{\omega}^{2}+1 \right) b}}-4\,{
\frac {M{\it A_0}}{ \left( {\it A_2}\,{\omega}^{2}+1 \right) ^{2}b}}
\beta+O \left( {\beta}^{2} \right) 
 \label{da1}
\end{equation}

The effect of the dark matter can be seen by comparison with the above deflection angle by Schwarzschild black hole. Hence dark matter gives small deflection angle compared to the standard case.

\section{Effect of medium on deflection angle of the Schwarzschild-like wormhole using the Gauss-Bonnet theorem }

In this section, we consider the static Schwarzschild-like wormhole
solution \cite{Damour:2007ap} with metric:

\begin{equation}
ds^{2}=-(1-2M/r+\lambda^{2})dt^{2}+\frac{dr^{2}}{1-2M/r}+r^{2}d\Omega_{(2)}^{2}\,,\label{wormhole}
\end{equation}
which reduces to the black hole metric in Eq. \ref{2} at $\lambda=0$. Using the transformation of $t\rightarrow t/\sqrt{1+\lambda^{2}}$
and $M\rightarrow M(1+\lambda^{2})$, the metric functions of the Schwarzschild-like wormhole spacetime becomes:
\begin{equation}
f(r)=1-\frac{2M}{r}\,,\quad g(r)^{-1}=1-\frac{2M(1+\lambda^{2})}{r}\,.
\end{equation}

\subsection{Case 1:} 

We first use the constant profile as refractive
$n_{m} =  n_0$ to calculate the deflection angle in the medium in weak field limits. 
Using the same procedure, we obtain the optical metric and calculate the Gaussian optical curvature for the Schwarzschild-like wormhole at linear order of M as follows:

\begin{equation}
K \approx -{\frac { \left( {\lambda}^{2}+2 \right) M}{{\rho}^{3}{{\it n_0}}^{2}}}+O(M^3),
\end{equation}
and after similar calculations, the corresponding deflection angle in the leading order terms is
\begin{equation}
\hat{\alpha}=2\,{\frac {M{\lambda}^{2}}{{\it n_0}\,b}}+4\,{\frac {M}{{\it n_0}\,b}}
\end{equation}
which agrees with the well-known results in the
limit in which the medium is negligible ($n_0=1$) \cite{Ovgun:2018fnk}.

\subsection{Case 2:}

To see the effect of the MFE-like medium, we use \cite{leonn}
\begin{equation}n = \frac{z_0}{1 + z^2}, \end{equation}
where $z_0$ and $z$ are a constant. The Gaussian curvature of the optical
metric approximating in leading orders is everywhere
negative and found as :

\begin{equation}
K \approx -{\frac { \left( {z}^{2}+1 \right) ^{2} \left( {\lambda}^{2}+2
 \right) M}{{\rho}^{3}{{\it z_0}}^{2}}}\end{equation}
Using the GBT, the  deflection angle is calculated as follows:
\begin{equation}
  \hat{\alpha} \simeq 2\,{\frac {M{\lambda}^{2}{z}^{2}}{{\it z_0}\,b}}+2\,{\frac {M{\lambda}^
{2}}{{\it z_0}\,b}}+4\,{\frac {M{z}^{2}}{{\it z_0}\,b}}+4\,{\frac {M}{{
\it z_0}\,b}}
 \hspace{0.17em}.
\end{equation}
At the $z =0$ and $z_0=1$,  it reduces to previous result \cite{Ovgun:2018fnk}.

\subsection{Case 3:}

Last we use the refractive index for the dark matter given in Eq. \ref{dm} to calculate the deflection angle of a wormhole in a medium. The Gaussian curvature is obtained as:
\begin{equation}
K \approx -{\frac { \left( {\lambda}^{2}+2 \right) M}{{\rho}^{3} \left( {\it A_2}
{\omega}^{2}+{\it A_0}\,\beta+1 \right) ^{2}}}
 \end{equation}
The deflection angle is found as follows:

\begin{equation}
\hat{\alpha}= 2\,{\frac {M{\lambda}^{2}}{ \left( {\it A_2}\,\,{\omega}^{2}+{
\it A_0}\,\beta+1 \right) b}}+4\,{\frac {M}{ \left( {\it A_2}\,
\,{\omega}^{2}+{\it A_0}\,\beta+1 \right) b}}  \label{da2} \end{equation}

We find that the deflected photon through the dark matter around the Schwarzschild-like wormhole has large deflection angle compared to the standard case.

\section{Conclusions} We have calculated the deflection angle of black holes and wormholes in a dark matter medium, using the GBT. This was achieved by constructing optical metrics. In summary, we have investigated that GBT is a partially topological effect. We have demonstrated this by using three different cases. In the first case, we used the constant profile as a refractive index. Then, by constructing the optical geometry and using the GBT, we have obtained  the deflection angle in the weak field limit. The deflection angle of the Schwarzschild black hole has been calculated correctly in a medium that has a constant $n_0$ refractive index. In the second case, we have used the MFE-like model (but uniform in large distances) and repeated the calculation and showed that it produces  a similar effect. 

In section 2, we have repeated our method on the Schwarzschild-like wormhole to see the effect of the dark matter medium when light is propagated through it. Note that we have supposed that refractive index is spatially non-uniform as long as it is uniform
at large distances. In the first case, we have again used the constant refractive index, and we have considered the MFE-like profile, and, finally, the medium for the dark matter is taken to find the deflection angle in the weak field limit. We have concluded that the deflection angle by black hole decreases in a medium of dark matter, as seen in Eq. (\ref{da1}), on the other hand  deflection angle by wormhole increases as seen in Eq. (\ref{da2}).

These results suggest that weak deflection within dark matter medium or MFE-like (perfect imaging), can be calculated using the Gibbons and Werner method which gives us hints to understand the nature of dark matter.  

\acknowledgments{A. \"{O}. is grateful to Institute for Advanced Study, Princeton for hospitality. This work was supported by Comisi{\'o}n Nacional de Ciencias y Tecnolog{\'i}a of Chile through FONDECYT Grant $N^\mathrm{o}$ 3170035 (A. {\"O}.)}










\end{document}